\documentclass[12pt]{article}
\pdfoutput=1
\usepackage{tikz}
\usetikzlibrary{matrix,arrows,decorations.pathmorphing,decorations.pathreplacing}
\usepackage{jheppub}
\usepackage{amsmath,amssymb,euscript,array,mathrsfs,appendix,ctable,marvosym}
\usepackage{arydshln}
\usepackage{todonotes}
\usepackage{graphicx}
\usepackage[normalem]{ulem}

\input{tcilatex}

\def\CF{{\cal F}}
\def\Tr{\text{Tr}}
\def\SU{\text{SU}}
\def\SO{\text{SO}}
\def\JJ{{\mathscr J}}
\newcommand{\EQ}[1]{\begin{equation}\begin{split} #1
\end{split}\end{equation}}

\title{S-Matrices and Quantum Group Symmetry of k-Deformed Sigma
Models}
\author[a]{Timothy J. Hollowood,}
\author[b]{J. Luis Miramontes}
\author[c]{and David M. Schmidtt}
\affiliation[a]{Department of Physics, Swansea University, Swansea, SA2 8PP, U.K.}
\affiliation[b]{Departamento de F\'\i sica de Part\'\i culas and IGFAE,
Universidad de Santiago de Compostela, 15782 Santiago de Compostela, Spain}
\affiliation[c]{Instituto de F\'\i sica Te\'orica IFT/UNESP, Rua Dr. Bento Teobaldo Ferraz 271, Bloco II, CEP 01140-070, S\~ao Paulo-SP, Brasil}

\emailAdd{t.hollowood@swansea.ac.uk}
\emailAdd{jluis.miramontes@usc.es}
\emailAdd{david.schmidtt@gmail.com}

\abstract{Recently, several kinds of integrable deformations of the string world sheet theory in the gauge/gravity correspondence have been constructed. One class of these, the $k$ deformations associated to the more general $q$ deformations but with $q=e^{i\pi/k}$ 
a root of unity, has been shown to be related to 
a particular discrete deformation of the principal chiral models and (semi-)symmetric space sigma models involving a 
gauged WZW model. We conjecture a form for the exact S-matrices of the bosonic integrable field theories of this type. The S-matrices imply that the theories have a hidden infinite dimensional affine quantum group symmetry.
We provide some evidence, via quantum inverse scattering techniques, that the theories do indeed
possess the finite-dimensional part of this quantum group symmetry.}

\notoc
\begin{document}

\maketitle

\newpage

\section{Introduction}

One of the most fascinating applications of the Quantum Inverse Scattering
Method (QISM) to sigma models has been to the exact quantization of the \SU(2)
Principal Chiral Model (PCM) in the 80's by Faddeev and
Reshetikhin (FR) in their seminal paper \cite{Faddeev-Reshetikhin}. The main conceptual
idea was to replace the non-ultra-local Poisson bracket of the theory by a
new ultra-local one that preserved the PCM equations of motion but under the
time flow of a new Hamiltonian. After doing this, it was possible to
construct a magnetic lattice algebra and to apply the QISM allowing for the 
the determination of the exact quantum spectrum and S-matrix of the \SU(2) PCM in a 
thermodynamic limit. However, one drawback of this new ultra-localization
trick was the lack of a systematic understanding, in the sense that the introduction
of the new Poisson bracket (to be quantized) was not rigorously justified
and the trick remained unused for many decades precluding the use of
the QISM to other sigma models beyond the $SU(2)$-PCM considered by FR.

This situation changed only very recently when Delduc, Magro and Vicedo
(DMV) in \cite{DMV 1}, using classical integrability techniques, succeeded in
identifying the algebraic structure behind the FR ultra-localization
mechanism. This allowed DMV to generalize the procedure to any PCM on a Lie
group $F$ and, more interestingly, to any sigma model on a symmetric space $F/G$, and also to the semi-symmetric spaces like $\text{AdS}_{5}\times S^{5}$ that appear in superstring theory and the \text{AdS}/CFT correspondence \cite{DMV 3}. The
generalized FR mechanism was considered by DMV only at classical level and it boils
down to a very simple recipe: it is equivalent to replacing in the Poisson
bracket the so-called twisting function $\varphi _{\sigma }(z)$ associated
to the sigma model by the new FR twisting function $\varphi _\text{FR}(z)$, which
is a simpler---in some well defined sense---function of the spectral parameter 
$z$. The next step in the implementation of the QISM is the introduction of
the lattice algebra regularization of the theory, which is the ultimate goal
of the quantization procedure. The lattice Poisson algebras
corresponding to the $F/G$ sigma models and to the $\text{AdS}_{5}\times S^{5}$
superstring were constructed explicitly in \cite{DMV 1,DMV 2} and, remarkably, they are deeply related to the Pohlmeyer
reduction of the sigma models \cite{Grigoriev,Andrei,Miramontes:2008wt}. Unfortunately,
the quantization of these lattice Poisson algebras is not known yet and it
is still an open problem.

Despite of the fact that the FR mechanism turned out to be a very simple
algebraic operation, a key ingredient was still missing, namely,
the replacement of the twisting function $\varphi _{\sigma }(z)\rightarrow
\varphi _\text{FR}(z)$ was not justified from a physical standpoint. After all,
changing the Poisson structure is equivalent to modifying the
kinetic term in the action functional and this raises the question as to whether it is
possible to find an integrable deformation of the original sigma models that
implements in a natural way the FR bracket in some limit of the
deformation parameter.

This led to the introduction of the $\eta $-deformed sigma models by DMV in 
\cite{eta 1,eta 2,derivation}, which are Yang-Baxter type
deformations of the form originally introduced by Klimcik in \cite{Klimcik}
for the PCM. The $\eta $-deformed Lagrangian combines the two twisting
functions $\varphi _{\sigma }(z),$ $\varphi _\text{FR}(z)$ into a new one $%
\varphi _{\epsilon (\eta )}(z)$ leading to a deformed interpolating Poisson
bracket of the form $\left\{ f,g\right\} _{\eta }=\left\{ f,g\right\}
_{\sigma }+\epsilon ^{2}(\eta )\left\{ f,g\right\} _\text{FR}.$ Unfortunately,
the reality condition of the action and integrability properties restrict the
domain of the parameter $\eta$. This  translates into $0\leq
\epsilon ^{2}(\eta )<1$, which means that the FR bracket is not reachable
in parameter space. In other words, it will never be the dominant bracket
that has to be quantized within the FR scheme.

An entirely different kind of deformation has been emerging based originally on the 
observation that the S-matrix that describes the scattering of excitations on the $AdS_5\times S^5$ string world sheet theory admits a 
quantum group deformation \cite{Hoare:2011nd,Hoare:2011wr,Hoare:2012fc,Hoare:2013ysa}. In turn, this was motivated by the observation that the world sheet equations of motion are equivalent, via the Polhmeyer reduction, to a relativistic generalized sine-Gordon theory \cite{Grigoriev,Andrei} and the deformed S-matrix provides an explicit relation between both theories at the quantum level. This deformation was recently identified in \cite{HMS fermionic}
as a deformation of the non-Abelian T-dual of the world sheet sigma
model related to the (bosonic) ``$\lambda$-deformations'' of the PCM introduced by Sfetsos in \cite{Sfetsos I} and of the symmetric space sigma models constructed by the present authors in \cite{HMS bosonic}. We call these theories the $k$-deformed sigma models, rather than $\lambda$-deformed, because the parameter $q=e^{i\pi/k}$ is fixed, a root of unity, while $\lambda$ is a coupling constant. In particular, in the bosonic models $\lambda$ is a running coupling in the quantum theory and, therefore, it is naturally interpreted
as the running coupling of the original sigma model. It is only in the classical limit $k\to\infty$ that $\lambda$ is a fixed parameter.

The $k$-deformed Lagrangian combines the two twisting functions and
Poisson brackets exactly as discussed above but with $-\infty <\epsilon
^{2}(\lambda )\leq 0,$ corresponding to $\lambda \subset \left[ 0,1\right] $,
so that in the limit $\lambda \rightarrow 0$ the action
functional indeed possesses the (generalized) FR bracket as its canonical Poisson
structure. Here, we will consider the bosonic models of this type.

Finding the lattice algebra regularization of the field theory is by far the
most difficult and important step toward the full implementation of the QISM.
It is equivalent to the regularization/quantization of the Maillet
bracket \cite{Maillet}\ associated to the field theory under examination, which, for the models of present interest, 
is non-ultra-local. The quantization of
the Maillet bracket is known to be a difficult and longstanding
unsolved problem for most of the interesting integrable field theories and
for this reason any successful attempt to regularize/quantize it, even in
some simplifying limit, has to be considered as a useful enterprise by itself. To date,
and in relation to the $k$-deformations, the (Maillet) lattice
algebra is only known in the limit $\lambda \rightarrow 0,$ at classical
level, $k=\infty$, and for any value of the spectral parameter $z$. The Poisson lattice
algebras coincide precisely with the ones introduced in the DMV papers
mentioned above.

In the present work, we find another limit in which we can regularize and
quantize the Maillet bracket. It is defined for any value of the 
parameter $\lambda \in \left[ 0,1\right] $ but only for a very specific
set of values of the spectral parameter that correspond to the poles of
the $\lambda $-deformed twisting function $\varphi _{\epsilon (\lambda )}(z)$%
. Remarkably, the lattice algebra turns out to be nothing but the Kac-Moody
lattice current algebra introduced in \cite{Unraveling,Faddeev}. In
particular, this allows us to show in a rather simple way that the $k$
-deformed sigma models associated to the principal chiral model on a Lie
group $F$ and to the bosonic sigma models on the symmetric spaces $F/G$,
both possess a finite-dimensional quantum group symmetry $U_{q}(
\mathfrak f)$ with a deformation parameter $q=e^{i\pi/k}$ being a root of unity. This
quantum group is the finite-dimensional piece of the
infinite-dimensional affine quantum group symmetry that the S-matrix of these
theories are conjectured to enjoy. Furthermore, we also show explicitly how these models naturally
implement the FR ultra-localization mechanism as a continuous limiting
process, i.e, as $\lambda \rightarrow 0$ we do have that $\left\{
f,g\right\} _{\lambda }\rightarrow \left\{ f,g\right\} _\text{FR}$.

It is important to emphasize that in this work we are considering the (bosonic) $k$-deformed theories as just $1+1$ dimensional integrable quantum field theories but not as string world sheet theories. Hence, we do not have the additional complications of a world-sheet theory, like gauge fixing and the Virasoro constraints, to contend with. In particular, in the situation studied here, the coupling $\lambda$ runs with scale, whereas in the stringy context we would expect $\lambda$ to be exactly marginal under the renormalization group. For applications of the $k$-deformation to string theory see \cite{HMS fermionic,Sfetsos:2014cea,Demulder:2015lva,Hoare:2014pna,Hoare:2015gda,Vicedo:2015pna,eta-lambda}.

The paper is organized as follows. In section \ref{s2}, we review the construction of the theories at the Lagrangian level. We then consider the theories at the quantum level and make a series of conjectures for the form of their exact S-matrices. This motivates the idea that the $k$-deformed theory has affine quantum group symmetries.
In section~\ref{s3}, we reveal the Poisson-Lie
group symmetry of the deformed sigma models, which is the
classical precursor of the quantum group. We write down the Maillet $\mathfrak r$/$\mathfrak s$
form of the Poisson brackets of the Lax operator, identify the twisting
functions and find their poles. We consider the deformed theories
and expand their monodromy matrix at the poles of the twisting function and
find the first non-trivial non-local charges suggesting the existence of an
affine quantum group. In section \ref{s4}, we review the lattice Kac-Moody algebra
and the quantization of the monodromy matrix evaluated at the poles of the
twisting function, which generates the finite-dimensional 
quantum group with a root-of-unity $q$-deformation parameter. Finally, we
draw some conclusions and directions for future work.

\section{The Deformed Sigma Models}\label{s2}

We will consider
two classes of bosonic $k$-deformed sigma models that we call the $k$-principal chiral models ($k$-PCM) and $k$-symmetric space sigma models ($k$-SSSM), associated to Lie groups $F$ and symmetric spaces $F/G$, respectively. They are constructed in the following way. For the $k$-PCM case \cite{Sfetsos I}, one takes the PCM for an $F$-valued field $f$ and then adds to it a WZW model for a second $F$-valued field $\CF$. The original PCM has an $F_L\times F_R$ global symmetry that acts as $f\to U_LfU_R^{-1}$. An interaction between the PCM and WZW model is generated by gauging the common $F$-action
\EQ{
f\to Uf\ ,\qquad \CF \to U\CF U^{-1}\ ,\qquad U\in F\ ,
\label{fact}
}
and fixing the gauge by setting $f=I$. The resulting theory has a global $F$ symmetry that one can view as arising from the original $F_R$-symmetry of the PCM.\footnote{It is worth pointing out, that this simple construction does not describe the $k$-deformation of the full superstring world sheet theory worked out in \cite{HMS fermionic}.}

For the $k$-SSSM case \cite{HMS bosonic}, one takes the $F/G$ sigma model defined in terms of an $F$-valued field $f$ and a gauge field $B_\mu$ corresponding to the $G$ right action $f\to fU^{-1}$ with $U\in G$. 
The $F/G$ sigma model also has a global $F_L$ symmetry that acts as $f\to U_L f$ with $U_L\in F$. Then, as for the
$k$-PCM case, one adds to it a WZW model for an $F$-valued field $\CF$ and gauges the common $F$-action \eqref{fact}. Finally, one partially fixes the gauge by setting $f=I$ and integrates out the gauge field $B_\mu$. This leaves the subgroup $G\subset F$ as the remnant gauge symmetry.

In both cases, the final theory has the form of a gauged WZW model
for a group $F$ where the whole group $F$, which acts by the anomaly-free vector action $\CF\to U\CF U^{-1}$, seems to be gauged. In fact, this is incorrect because the action involves a deformation that breaks the gauge symmetry completely for the $k$-PCM case, and breaks it down to the subgroup $G\subset F$ for the $k$-SSSM case:\footnote{In the Hamiltonian formulation, the first class constraints associated to $F$ are transformed into second class constraints for the $k$-PCM case. For the $k$-SSSM this fate also befalls the first class constraints in the coset part of the Lie algebra $\mathfrak f^{(1)}$ to leave a $G$ gauge symmetry.}
\EQ{
S_q&=-\frac k{2\pi}\int d^2x\Tr\Big[
\CF ^{-1}\partial_+\CF \,\CF ^{-1}\partial_-\CF +2A_+\partial_-\CF \CF ^{-1}\\ &~~~~~~~~~
-2A_-\CF ^{-1}\partial_+\CF -2\CF ^{-1}A_+\CF  A_-+2A_+\Omega\, A_-\Big]
\\ &~~~~~~~~~+\frac k{12\pi}\int d^3x\,\epsilon^{abc}\Tr\,\Big[\CF ^{-1}\partial_a\CF \,
\CF ^{-1}\partial_b\CF \,\CF ^{-1}\partial_c\CF \Big]\ .
\label{gWZW}
}
This action has two parameters: the level $k$, which is a positive integer that defines the deformation parameter
$q=\exp[i\pi/(c_2(F)+k)]$, and the
coupling constant $\lambda$ which is related to the coupling constant $\kappa$ of the original sigma models by means of $\lambda ^{-1}=1+\kappa ^{2}/k$.

The deformation appears  via the term that depends on the quantity $\Omega$ which takes the form\footnote{Notice that $\Tr(A_+\,\Omega A_-)=\Tr(\Omega A_+\,A_-)$ so that, in these bosonic models, $\Omega$ coincides with its transpose.}
\begin{equation}
\Omega=\begin{cases}\lambda ^{-1} & k\text{-PCM}\\ 
\mathbb P^{(0)}+\lambda^{-1}\mathbb P^{(1)} &
k\text{-SSSM}\ .\end{cases}
\end{equation}%
In the latter case, the $\mathbb P^{(i)}$, $i=0,1$, are the projectors along the $\mathbb{Z}
_{2}$ decomposition of the Lie algebra $\mathfrak{f=f}^{(0)}%
\mathfrak{\oplus f}^{(1)}$, where $\mathfrak{f}^{(0)}$ is the Lie algebra of 
$G$. Note that the $\mathbb Z_2$ gradation here is one of the defining elements of a symmetric space. 

In \eqref{gWZW},
the status of the field $A_\mu$ is rather subtle. In the $k$-PCM case, it is an auxiliary field that can be integrated out since it appears quadratically in the action. In the $k$-SSSM case, the component $A_\mu^{(1)}$ is, similarly, an auxiliary field, but the component $A_\mu^{(0)}$ is a genuine gauge field that reflects a $G$-gauge symmetry of the theory. It can be integrated out, but then one would have to fix the gauge on the field $\CF$. Alternatively one can fix the gauge by imposing, for example, $A_+^{(0)}=0$.

In the limit $\lambda\to1$, with $k\to\infty$ in such a way that 
\EQ{
k\Big(\lambda^{-1}-1\Big)=\kappa^2
}
is fixed, the theories can then be shown to be the non-abelian T-duals of the PCM for the group $F$, or of the symmetric space sigma model for $F/G$, respectively, with coupling constant $\kappa$. Theories which are non-abelian T-duals of each other are related by a canonical transformation and therefore it is a reasonable hypothesis that they can be equivalent at the quantum level.
The $k$-PCM has a global $F$ symmetry which acts as $\CF\to U\CF U^{-1}$ which is the remnant of the original $F_R$ symmetry of the PCM, while the $k$-SSSM has no obvious symmetries.

\subsection{Integrability}

The $k$-deformed theories are integrable at the classical level.\footnote{Proved originally for the $k$-PCM case in \cite{Sfetsos I} and more generally and in a simpler way in \cite{HMS bosonic}.}
To see this, we can solve for the auxiliary field $A_\mu$ in the form
\begin{equation}
A_{+}=\left( \Omega-\text{Ad}_{\CF^{-1}}\right) ^{-1}\mathcal{F}^{-1}\partial _{+}%
\mathcal{F},\text{ \ \ \ \ \ }A_{-}=-\left( \Omega -\text{Ad}_{\CF}\right) ^{-1}\partial
_{-}\mathcal{FF}^{-1}\ .  \label{connections}
\end{equation}%
The equations-of-motion can then be written in Lax form:
\EQ{
[\partial_++{\mathscr L}_+(z),\partial_-+{\mathscr L}_-(z)]=0\ ,
\label{zcc}
}
where, for the $k$-PCM 
\EQ{
{\mathscr L}_{\pm }(z) =\frac{z}{z\pm 1}\cdot\frac{2}{(1+\lambda )}A_{\pm }\ ,  \label{PCM Lax}
}
while for the $k$-SSSM
\EQ{ 
{\mathscr L}_{\pm }(z) =A_{\pm }^{(0)}+z^{\pm 1}\lambda^{-1/2}A_{\pm }^{(1)}\ .\label{Sigma Lax}
}

\subsection{Quantization and S-matrices}
\label{s2.2}

For integrable theories, one way to quantize the theories directly is to intuit a form for the exact S-matrix by using a combination of the global symmetries, the standard S-matrix axioms along with factorization and the Yang-Baxter equation that result from integrability. Integrability at the quantum level is thought to follow if there are higher spin conserved, possibly non-local, charges whose conservation is non-anomalous in the quantum theory. 
The integrability of the PCM is known to be non-anomalous for any group $F$, while the absence of an anomaly for the $F/G$ SSSM is not guaranteed in general. For instance, for compact symmetric spaces of definite signature it requires that the subgroup $G$ is simple \cite{Abdalla:1982yd}.\footnote{Or requires the addition of supersymmetry.} The quantum integrable SSSMs are then associated to either Type II symmetric spaces of the form $G\times G/G$, which provide an equivalent way to formulate the PCM for the group $G$, or Type I symmetric spaces of the form
\EQ{
&\SU(N)/\text{SO}(N)\ ,~~~\qquad\SU(2N)/\text{Sp}(N)\ ,\\
&\SU(2)/\text{U}(1)\equiv\mathbb CP^1\ ,\quad
\text{SO}(N)/\text{SO}(N-1)\equiv S^{N-1}\ ,
}
plus a number of possibilities involving the exceptional groups that we shall not consider here.

We make the hypothesis that the 
$k$-deformation of these theories are also integrable at the quantum level. One way 
to motivate this is to find a suitable ansatz for their S-matrices which can then be subject to various consistency tests. Note that there are effectively two different kinds of $k$-deformations of the PCM because they can also be formulated as a SSSM.

The first issue to consider is how the couplings $k$ and $\lambda$ flow in the quantum theory. The integer level $k$, of course, cannot have non-trivial RG flow because it multiplies the topological Wess-Zumino term in the action. However,
the coupling $\lambda$ does have a non-trivial RG flow. The loop counting parameter is $1/k$ and to the one-loop level the beta function is\footnote{For the $k$-PCM case the beta function can be extracted from the general approaches in
\cite{Tseytlin:1993hm,Itsios:2014lca,Sfetsos:2014jfa}. For the $k$-SSSM case, the beta function has been calculated in \cite{NEW-Tim}.}
\EQ{
k\text{-PCM:}&\qquad\mu\frac{d\lambda}{d\mu}=-\frac{2c_2(F)}{k}\Big(\frac\lambda{1+\lambda}\Big)^2\ ,\\
k\text{-SSSM:}&\qquad
\mu\frac{d\lambda}{d\mu}=-\frac{c_2(F)}{k}\lambda\ .
\label{bft}
}
So in both cases, the coupling $\lambda$ runs to zero in the UV.
In this limit, by integrating out the auxiliary field $A_\mu$, for the PCM case, and $A_\mu^{(1)}$, for the SSSM case, the theories take the form of a WZW model deformed by a current-current operator. For the PCM this is an ordinary (i.e.~non-gauged) WZW model
\EQ{
S_\lambda=S_\text{WZW}[\CF]+\frac{4\pi}{k\lambda}\int d^2x\,\Tr\big(\hat\JJ_+\hat\JJ_-\big)+\cdots\ ,
\label{wee1}
}
where $\hat\JJ_\pm$ are the standard Kac-Moody currents of the WZW model:
\EQ{
\hat\JJ_+=-\frac k{2\pi}\CF^{-1}\partial_+\CF\ ,\qquad\hat\JJ_-=\frac k{2\pi}\partial_-\CF\CF^{-1}\ .
}
In this case the deformation is marginally relevant, as is also clear from the form of the beta function \eqref{bft}.

For the SSSM case, the UV theory is a deformation of the gauged WZW for $F/G$,
\EQ{
S_\lambda=S_\text{gWZW}[\CF,A_\mu^{(0)}]+\frac{4\pi}{k\lambda}\int d^2x\,\text{Tr}\,\big(\hat\JJ_+^{(1)}\hat\JJ_-^{(1)}\big)+\cdots\ ,
\label{wee2}
}
where $\hat\JJ_\pm$ are the Kac-Moody currents of the $F/G$ gauged WZW model
\EQ{
\hat\JJ_+&=-\frac k{2\pi}\big(\CF^{-1}\partial_+\CF+\CF^{-1}A_+^{(0)}\CF-A_-^{(0)}\big)\ ,\\
\hat\JJ_-&=\frac k{2\pi}\big(\partial_-\CF\CF^{-1}-\CF A_-^{(0)}\CF^{-1}+A_+^{(0)}\big)\ .
}
Notice that the deformation now is relevant.
In the case of a Type II symmetric space, it is worth pointing out that the WZW model involves the product group $F=G\times G$ and so there are two independent levels $k_1$ and $k_2$.

The integrable deformations of WZW models, non-gauged  in \eqref{wee1} and gauged in \eqref{wee2} have been studied from a CFT perspective in \cite{Ahn:1990gn}.

There is one example of the $k$-deformed sigma models that has been studied in some detail and will provide the inspiration for the other cases. This corresponds to the PCM case with $F=\SU(2)$ \cite{Evans:1994hi}. To start with, the PCM models associated to the classical groups have a factorized S-matrix that can be written schematically in terms of 
 2 particle to 2-particle matrix elements of the form \cite{Abdalla:1984iq,Abdalla:1985ds}
\EQ{
S^{ab}(\theta)=X^{ab}(\theta)\widetilde S^{ab}_{(\infty)}(\theta)\otimes\widetilde S^{ab}_{(\infty)}(\theta)\ ,
\label{yhs}
}
where $\theta=\theta_1-\theta_2$ is the rapidity difference of the incoming states. The $X^{ab}(\theta)$ are scalar factors, known also as also dressing phases, that are needed to satisfy the S-matrix constraints of unitarity and crossing symmetry. The states of the theory are labelled by $a$ and $b$ which range over $1,2,\ldots,r$, the rank of $F$. They
transform in a product of representations of $F$, $W_a\otimes W_a$, where $W_a$ is the $a^\text{th}$ fundamental, or antisymmetric, representation, for unitary and symplectic groups, but are not irreducible for the orthogonal cases \footnote{If $V_a$ are the fundamental representations, then $W_a=\oplus_{j=0}^{a-2j\geq0}V_{a-2j}$.}. The two factors reflect the 
$F_L\times F_R$ symmetry of the PCM which act on the group field via the left and right action: $f\to U_LfU^{-1}_R$. In \eqref{yhs}, the $\widetilde S^{ab}_{(\infty)}$ are intertwiners
\EQ{
\widetilde S^{ab}_{(\infty)}:\quad W_a(\theta_1)\otimes W_b(\theta_2)\rightarrow W_b(\theta_2)\otimes W_a(\theta_1)\ .
}
These intertwiners are the rational solutions of the Yang-Baxter equation. The significance of the $\infty$ label will emerge.
The S-matrix is covariant under the $F_L\times F_R$ symmetry of the theory, but this symmetry extends to an affine Yangian symmetry. The full spectrum of the theory can then be obtained by using the bootstrap equations. The bound states are associated to the antisymmetric representations of $F$ although for the $\SO(N)$ groups these representations are sometimes reducible. 

For $F=\SU(2)$, in~\cite{Evans:1994hi} it was conjectured a form for the $S$-matrix of the $k$-deformed theory of the form
\EQ{
S(\theta)=X_{(k,\infty)}(\theta)\widetilde S_{(k)}(\theta)\otimes\widetilde S_{(\infty)}(\theta)\ ,
}
where the $\SU(2)_L$ factor, the one that is gauged in order to construct the deformed theory, has changed from the rational
$\SU(2)$ factor to the quantum group generalization with a quantum  group deformation parameter 
that is a root of unity: $q^{2(k+2)}=1$.
It is also important that
the quantum group factor $\widetilde S_{(k)}(\theta)$ is taken in the so-called interaction-round-a-face" (IRF) picture, or ``restricted-solid-on-solid" (RSOS) picture. The significance of this is that the quantum numbers in this sector are naturally interpreted as kinks. So states of the theory carry a kink charge $\pm1$ and also a vector in the two-dimensional representation of $\SU(2)$. The detailed form of the S-matrix is written in \cite{Evans:1994hi}. In the present context, the fact that the particle states are kinks can be understood as arising form the non-trivial spatial boundary conditions of the WZW models.\footnote{Discussed, for example, in \cite{NA Kinks}.}

Based on the understanding of the $\SU(2)$ $k$-PCM model above,
it is natural to conjecture that the $k$-PCM for a generic group $F$ have a S-matrix of the schematic form
\EQ{
S^{ab}(\theta)=X^{ab}_{(k,\infty)}(\theta)\widetilde S^{ab}_{(k)}(\theta)\otimes\widetilde S^{ab}_{(\infty)}(\theta)\ ,
\label{pmm}
}
with $q^{2(k+c_2(F))}=1$, where $c_2(F)$ is the dual Coxeter number of $F$. 
As in the $\SU(2)$ case, the quantum group intertwiner $\widetilde S^{ab}_{(k)}(\theta)$ is defined in the IRF or RSOS representation. These S-matrices in general were first written down in \cite{Ahn:1990gn} and discussed in detail 
in \cite{Hollowood:1993fj}. Note that the scalar factors $X^{(ab}_{(k,\infty)}$ here, and below, depend implicitly on the actual group $F$.
Using TBA techniques, this conjecture has been checked for $SU(2)$ in~\cite{Evans:1994hi} and, more recently, for $SU(N)$ in \cite{NEW-Tim}. 

The intertwiners $\widetilde S^{ab}_{(k)}(\theta)$ are invariant under the action of an affine quantum group $U_q(\hat{\mathfrak f})$ in the following sense. There is an associated co-product $\Delta$ which describes how the generators act on a tensor product:
$\Delta(u)W_a(\theta_1)\otimes W_b(\theta_2)$, $u\in U_q(\hat{\mathfrak f})$.
The theory is invariant under the affine quantum symmetry in the sense that 
\EQ{
\Delta(u)S^{ab}(\theta_{12})=
S^{ab}(\theta_{12})\Delta(u)\ , \qquad u\in U_q(\hat{\mathfrak f})\ .
}
So the S-matrix elements lie in the commutant of $U_q(\hat{\mathfrak f})$ acting on a tensor product representation. In the limit, $k\to\infty$, the quantum group reduces to the Yangian symmetry of the undeformed PCM.

Now we turn to the SSSM. For the Type II cases, there is an obvious form that the S-matrix should take. One simply generalizes \eqref{pmm} to
\EQ{
S^{ab}(\theta)=X_{(k_1,k_2)}^{ab}(\theta)\widetilde S^{ab}_{(k_1)}(\theta)\otimes\widetilde S^{ab}_{(k_2)}(\theta)\ ,
\label{pmm2}
}
which means that the symmetry involves a tensor product of quantum groups $U_{q_1}(\hat{\mathfrak f})\otimes 
U_{q_2}(\hat{\mathfrak f})$.

For the Type I theories, a bespoke construction is needed in each case based on the S-matrices that were constructed for the undeformed theories \cite{Abdalla:1986xb}. However, the form of the undeformed S-matrix leads to a natural conjecture in each case.

\noindent
\underline{$S^{N-1}$ (including $\mathbb CP^1\equiv S^2$)}: for this series the S-matrix has the form 
originally found by Zamolodchikov and Zamolodchikov
\cite{Zamolodchikov:1978xm}
\EQ{
S(\theta)=X_{(\infty)}(\theta)\widetilde S_{(\infty)}(\theta)\Big|_{\SO(N)}\ ,
\label{pmm3}
}
with particles transforming on the $N$-dimensional defining representation of $F=SO(N)$, which we show explicitly. We conjecture that the $k$-deformed theory has an S-matrix of the form
\EQ{
S(\theta)=X_{(k)}(\theta)\widetilde S_{(k)}(\theta)\Big|_{\SO(N)}\ ,
\label{pmm4}
}
with $q^{2(N-2+k)}=1$.

\noindent
\underline{$\SU(N)/\text{SO}(N)$}: for this series the S-matrix has the form
\EQ{
S(\theta)=X_{(\infty)}(\theta)\widetilde S_{(\infty)}(\theta)\Big|_{\SU(N)}\otimes_\text{s} \widetilde S_{(\infty)}(\theta)\Big|_{\SU(N)}\ ,
\label{pmm5}
}
where $\otimes_\text{s}$ indicates the symmetric product and the 
particles transform in the symmetric $N(N+1)/2$-dimensional representation of $F=SU(N)$ that acts on both factors. In this case, there are no bound states. There is an analogous set of anti-particles transforming in the conjugate representation \cite{Abdalla:1985ds}. 
We conjecture that the $k$-deformed theory has an S-matrix of the form
\EQ{
S(\theta)=X_{(k)}(\theta)\widetilde S_{(k)}(\theta)\Big|_{\SU(N)}\otimes_\text{s} \widetilde S_{(k)}(\theta)\Big|_{\SU(N)}\ ,
\label{pmm6}
}
with $q^{2(N+k)}=1$.

\noindent
\underline{$\SU(2N)/\text{Sp}(N)$}: for this series the S-matrix has the form
\EQ{
S^{ab}(\theta)=X^{ab}_{(\infty)}(\theta)\widetilde S_{(\infty)}(\theta)\Big|_{\SU(2N)}\otimes_\text{a}\widetilde S_{(\infty)}(\theta)\Big|_{\SU(2N)}\ ,
\label{pmm7}
}
where $\otimes_a$ indicates the antisymmetric product. For each $a$, the particles transforms in a product of  fundamental
representations $V_a\otimes V_a$, one for each factor. Then the $\otimes_\text{a}$ indicates that these are projected onto the anti-symmetric term of the $\SU(2N)$ which acts diagonally on both factors.
 We conjecture that the $k$-deformed theory has an S-matrix of the form
\EQ{
S^{ab}(\theta)=X^{ab}_{(k)}(\theta)\widetilde S_{(k)}(\theta)\Big|_{\SU(2N)}\otimes_\text{a}\widetilde S_{(k)}(\theta)\Big|_{\SU(2N)}\ ,
\label{pmm8}
}
with $q^{2(2N+k)}=1$.

\section{Poisson-Lie group symmetry}\label{s3}

In this section, we will look for the classical footprints of a quantum
group which appear in the form of global Poisson-Lie group symmetries of the
Lagrangian. We will show that the conserved Poisson-Lie charges are given by
the monodromies of two (mutually commuting) Kac-Moody currents.

\subsection{Non-local symmetries of the $\protect\lambda $-deformed actions}

The action (\ref{gWZW}) is invariant under the
discrete ${\mathbb Z}_2$-transformation\footnote{Since the action is also invariant under $(\CF,A_\pm,x)\to(\CF^{-1},A_\mp,-x)$, this transformation is equivalent to the one written down in~\cite{Hoare:2015gda}.}
\EQ{
\Pi(\CF,A_+,A_-,\lambda,k)=\big(\CF^{-1}, \text{Ad}_{\CF^{-1}} (A_+ + \partial_+\CF\CF^{-1}), \Omega A_-, \lambda^{-1}, -k\big)
}
which, in particular, allows one to extend the range of the parameter $\lambda$ from $[0,1]$ to $[0,\infty)$. Then, after integrating out the fields $A_\pm$, the resulting effective action turns out to be invariant under 
\EQ{
\Pi(\CF,\lambda,k)=(\CF^{-1},\lambda^{-1},-k)
\label{parity}
}
as pointed out originally in~\cite{Itsios:2014lca}, where it was shown that
it severely restricts the form of the RG
flow equation for $\lambda$. This last symmetry will also play an important role in the following.

We now show that the theories have global Poisson-Lie group symmetries.
To do this, the natural starting point is to consider left and right infinitesimal $F$-transformations separately. Namely,
\EQ{
&\delta_L\CF= \epsilon_L \CF\,,\qquad \delta A_+= [\epsilon_L,A_+]-\partial_+\epsilon_L\,,\qquad \delta A_-=0\,,\\[5pt]
&\delta_R\CF= \epsilon_R \CF\,,\qquad \delta A_+=0\,,\qquad \delta A_-= [\epsilon_R,A_-]-\partial_+\epsilon_R
}
with $\epsilon_{L},\epsilon_R\in{\mathfrak f}$.
We find%
\EQ{
\delta _{L}S&=-\frac{k}{\pi }\dint d^{2}x\,\text{Tr}\left( \epsilon _{L}\left[
\partial _{+}+A_{+},\partial _{-}+\Omega A_{-}\right] \right)\ ,\\
\delta _{R}S&=-\frac{k}{\pi }\dint d^{2}x\,\text{Tr}\left( \epsilon _{R}\left[
\partial _{+}+\Omega A_{+},\partial _{-}+A_{-}\right] \right)\ .
\label{left-right variations}
}

In the $k$-PCM case, the vector action of the group corresponds to taking
 $\epsilon _{R}=-\epsilon $, $\epsilon _{L}=\epsilon $, and then adding both
variations. This leads to%
\begin{equation}
\delta _{V}S=-\frac{k(1-\lambda)}{\pi \lambda } \dint
d^{2}x\,\text{Tr}\,\left(\epsilon\left( \partial _{+} A_{-}+\partial _{-}
A_{+}\right)\right)
=-\frac{k(1-\lambda)}{2\pi \lambda } \dint
d^{2}x\,\text{Tr}\,\left(\epsilon\, \partial _\mu A^\mu\right)\ .
\label{vector charge}
\end{equation}%
Then, for global transformations, $\delta _{V}S=0$ provided that the fields $A_\pm$ vanish at infinity. This shows that the action is invariant off-shell 
under global vector transformations, as expected, and provides conserved Noether charges. As a consequence, the target space
geometry of the deformed PCM is invariant under the vector action of the
group $F$.
In the $k$-SSSM case, we get instead
\EQ{
\delta_V S=-\frac{k(1-\lambda)}{\pi \lambda } \dint
d^{2}x\,\text{Tr}\, \left\{\epsilon\, \left([\partial_+ +A_+^{(0)}, A_-^{(1)}] + [\partial_- +A_-^{(0)}, A_+^{(1)}]\right)\right\}\,,
\label{noNoether}
}
which vanishes off-shell for $\epsilon$ in the Lie algebra of $G$. This exhibits the invariance of the action under $G$ gauge transformations, but also that the $k$-SSSM has no other obvious off-shell (Noether) symmetries. However, since these theories are classically integrable, this is not the end of the story.

Going back to the separate transformations, \eqref{left-right variations}, we note that both can be put in the form
\begin{equation}
\delta _{LP}S\sim \dint \epsilon ^{a}\Big\{ dJ^{a}+\frac{1}{2}\widetilde{f}_{%
\text{ \ }bc}^{a}J^{b}\wedge J^{c}\Big\}\ ,
\end{equation}%
for some current $J$ and some structure constants $\widetilde{f}_{\text{ \ }%
bc}^{a}.$ On shell, each current is flat and so there exists conserved quantities $\Gamma$, the non-abelian moments, such that $J=\Gamma ^{-1}d\Gamma$. It is well-known that the existence of conserved non-local charges such as these have played an important role proving the integrability of sigma models and finding their exact S-matrices; for example in 
\cite{Abdalla:1985ds}.

To be more explicit, let us recall that the invariance of the action under generic field transformations provides the equations-of-motion which, in our case, can always be written in the Lax form~\eqref{zcc}. Then, the monodromy matrix\footnote{We assume here\sout{,} that the field theories are defined on a spatial circle.}
\begin{equation}
m(z)=\text{Pexp} \left[ \dint_{S^{1}}dx\,{\mathscr L}(x,z)\right]\,,\qquad
{\mathscr L}= -{\mathscr L}_+  + {\mathscr L}_-\,,
\label{Monodromy}
\end{equation}%
is conserved for generic values of $z$, which
provides an infinite number of, generally non-local, conserved quantities.
Remarkably, the left and right variations~\eqref{left-right variations} 
can be written in terms of
the Lax pair evaluated at specific values of the spectral parameter~$z$:%
\EQ{
\delta _{L}S &=-\frac{k}{\pi }\dint d^{2}x\,\text{Tr}\left( \epsilon _{L}\left[
\partial _{+}+{\mathscr L}_{+}(z_{-}),\partial _{-}+{\mathscr L}_{-}(z_{-})%
\right] \right) ,  \\
\delta _{R}S &=-\frac{k}{\pi }\dint d^{2}x\,\text{Tr}\left( \epsilon _{R}\left[
\partial _{+}+{\mathscr L}_{+}(z_{+}),\partial _{-}+{\mathscr L}_{-}(z_{+})%
\right] \right) ,  
 \label{variations}
 }
where in the $k$-PCM case
\EQ{
z_{\pm }=\pm \frac{\lambda +1}{\lambda -1}\ ,
\label{evaluation points}
}
and in the $k$-SSSM case
\EQ{
z_{+}^{2}=\lambda ^{-1}\ ,\qquad z_{-}^{2}=\lambda\ .
\label{evaluation points sigma}
}
In both, the significance of $z_\pm$ is that
\EQ{
{\mathscr L}_{+}(z_{+})=\Omega A_{+}\ ,\quad{\mathscr L}_{-}(z_{+})=A_{-}\ ,\quad {\mathscr L}_{+}(z_{-})=A_{+}\ ,\quad{\mathscr L}_{-}(z_{-})=\Omega A_{-}\ .
}
The equations~\eqref{variations} relate the conservation of $m(z_+)$ and $m(z_-)$ to the invariance of the action under left and right transformations, respectively, in the following sense. $\delta _{L/R}S=0$ implies that ${\mathscr L}_{\pm}(z_{\mp})$ is flat, which provides the conserved quantities $m(z_\mp)$. 

Notice that the special points $z_{\pm}$ are uniquely determined by the
deformation parameter $\lambda $ and, furthermore, they are 
mapped into each other under the discrete transformation \eqref{parity}. 
Below, we will see that these points are precisely the poles of the so-called twisting functions that underlies the Poisson structure of these theories.

The Hamiltonian structure of the $k$-PCM and the $k$-SSSM can be written in terms of the canonical currents~\cite{HMS bosonic}
\EQ{
\mathscr J_{+}&=-\frac{k}{2\pi }\left( \mathcal{F}^{-1}\partial _{+}\mathcal{
F}+\mathcal{F}^{-1}A_{+}\mathcal{F}-A_{-}\right)\ ,\\ \mathscr{
J}_{-}&=\frac{k}{2\pi }\left( \partial _{-}\mathcal{FF}^{-1}-\mathcal{F}A_{-}
\mathcal{F}^{-1}+A_{+}\right) \ , \label{KM currents}
}
which obey the classical Kac-Moody algebra\footnote{Here, we define $\delta_{xy}\equiv\delta(x-y)$ and follow the conventions of \cite{HMS bosonic}.}
\begin{equation}
\left\{ \mathscr J_{\pm }^{a}(x),\mathscr J_{\pm }^{b}(y)\right\} =f^{abc}%
\mathscr J_{\pm }^{c}(x)\delta _{xy}\pm \frac{k}{2\pi }\delta ^{ab}\delta
_{xy}^{\prime }\ ,\quad \left\{ \mathscr J_{+}^{a}(x),\mathscr J%
_{-}^{b}(y)\right\} =0.  \label{KM algebra}
\end{equation}%
On-shell, we can write these currents as%
\begin{equation}
\mathscr J_{+}=-\frac{k}{2\pi }\left( \Omega A_{+}-A_{-}\right) ,\text{
\ \ }\mathscr J_{-}=-\frac{k}{2\pi }\left( \Omega A_{-}-A_{+}\right) .
\label{onshellKM}
\end{equation}%
The key observation for the following analysis if that the Kac-Moody currents are related to the Lax connection in a very simple way; namely,
\begin{equation}
{\mathscr L}(z_{+})=\frac{2\pi }{k}\mathscr J_{+},\text{ \ \ \ \ \ }\mathscr{%
L}(z_{-})=-\frac{2\pi }{k}\mathscr J_{-}\ ,  \label{evaluation}
\end{equation}%
where ${\mathscr L}=-{\mathscr L}_++{\mathscr L}_-$ is the spatial component of the Lax connection.
This simple observation will allow us to deduce below that the $k$-deformed
field theories have a quantum group symmetry. Then,
from~\eqref{evaluation} we find that the conserved quantities
associated to the global $F_{L}\times F_{R}$ Poisson-Lie group action in
both models take a rather universal form
\begin{equation}
\ m(z_{+})=\text{Pexp} \left[ \frac{2\pi }{k}\dint_{S^{1}}dx\ \mathscr J_{+}(x)%
\right] ,\text{ \ \ \ \ \ }m(z_{-})=\text{Pexp} \left[ -\frac{2\pi }{k}%
\dint_{S^{1}}dx\,\mathscr J_{-}(x)\right]\ .
 \label{non-Abelian moments}
\end{equation}%
Applying \eqref{parity} to either~\eqref{KM currents} or~\eqref{onshellKM} we see that the effect of the discrete
transformation is to exchange the two Kac-Moody currents 
\begin{equation}
\Pi \mathscr J_{\pm }=\mathscr J_{\mp }\ .
\end{equation}
Then, $m(z_\pm)$ are mapped into each
other under~\eqref{parity}, so that one of the Poisson-Lie
symmetries, say $F_R$, can be generated by a combination of $F_{L}$ and $\Pi $. In section~\ref{s4} below we turn to the  quantum version of the algebra generated by 
$m(z_{\pm })$.

\subsection{The $\mathfrak r$/$\mathfrak s$ Maillet forms and twisting functions}

In this section, we exploit the Kac-Moody algebraic structure (\ref{KM
algebra}) of the $k$-deformed theories to compute the Poisson brackets of the spatial
component of the Lax connection. In particular, we are interested in
extracting the twisting function $\varphi _{\epsilon }(z)$ induced by the $%
\mathfrak r$/$\mathfrak s$ form and study its pole structure. We begin with the $k$-PCM theories, and then consider the $k$-SSSM theories, which require a slightly different treatment.

\noindent{\bf\underline{ $k$-PCM case}:} 

\noindent
We start by writing the spatial component of the $k$-PCM Lax connection for generic values of the spectral parameter in the form 
\begin{equation}
{\mathscr L}(z)=f_{-}(z)\mathscr J_{+}+f_{+}(z)\mathscr J_{-}\ ,
\label{spatial Lax}
\end{equation}%
with
\begin{equation}
f_{\pm }(z)=-\frac{z}{1-z^{2}}\left( p\pm qz\right)\ ,\text{ \ \ }p=-\frac{%
4\pi \lambda }{k(1-\lambda ^{2})}\ ,\text{ \ \ }q=-\frac{4\pi \lambda }{%
k(1+\lambda )^{2}}\,.
\end{equation}%
Then, the Kac-Moody algebra~\eqref{KM algebra} provides the Maillet $\mathfrak r$/$\mathfrak s$ representation of the Poisson bracket
\EQ{
\big\{\overset1{\mathscr L}(x,z),\overset2{\mathscr L}(y,w)\big\} &=\big[ \mathfrak r,%
\overset1{\mathscr L}(x,z)+\overset2{\mathscr L}(y,w)\big] \delta _{xy}\\ &\qquad-\big[ \mathfrak s,
\overset1{\mathscr L}(x,z)-\overset2{\mathscr L}(y,w)\big] \delta _{xy}-2\mathfrak s\delta
_{xy}^{\prime }\ .  \label{Lax-Lax PB}
}
In this kind of representation, the equation is to be understood as acting on some tensor product $V\otimes V$, where $V$ is some faithful representation of the Lie algebra $\mathfrak f$, and the index on top show which of the factors the quantity acts on, e.g.~$\overset1{\mathscr L}(V\otimes V)=(\mathscr LV)\otimes V$ and 
$\overset2{\mathscr L}(V\otimes V)=V\otimes(\mathscr LV)$, etc. Quantities like $\mathfrak s$ and $\mathfrak r$ defined below act on both factors.

In \eqref{Lax-Lax PB}, the $\mathfrak r$/$\mathfrak s$ matrices are given by
\EQ{
\mathfrak r(z,w) =\left[ \frac{w\varphi _{\epsilon }(z)^{-1}+z\varphi
_{\epsilon }(w)^{-1}}{z-w}\right]\mathfrak C\ ,  \quad
\mathfrak s(z,w) =\left[ \frac{w\varphi _{\epsilon }(z)^{-1}-z\varphi
_{\epsilon }(w)^{-1}}{z-w}\right]\mathfrak C\ ,  
\label{r/s matrices}
}
where 
\begin{equation}
\varphi _{\epsilon }(z)=2p\cdot \frac{1}{z}\cdot \frac{
1-z^{2}}{q^{2}z^{2}-p^{2}}
\end{equation}%
is the so-called twisting function and $\mathfrak C=\sum_aT^{a}\otimes T^{a}$ is the
tensor Casimir. Notice that $\varphi _{\epsilon }(z)$ has two poles located precisely at
at $z_{\pm }$ 
\begin{equation}
z_{\pm }=\mp \frac{p}{q}=\pm \frac{\lambda +1}{\lambda
-1}\ .
\end{equation}%
These poles are exactly the evaluation points displayed in
(\ref{evaluation points}) used to extract the Lie-Poisson group charges out of the
monodromy matrix. The extra pole at $z=0$ has no physical significance and can be removed
by working with $z\rightarrow 1/z$ instead. In this case, the poles are inverted $z_{\pm
}\rightarrow z_{\pm }^{-1}$ and the (\ref{r/s matrices}) can be rewritten in
a more suggestive way%
\EQ{
\mathfrak r(z,w) =-\left[ \frac{\phi _{\epsilon }(z)^{-1}+\phi _{\epsilon
}(w)^{-1}}{z-w}\right] \mathfrak C\ , \quad
\mathfrak s(z,w) =-\left[ \frac{\phi _{\epsilon }(z)^{-1}-\phi _{\epsilon
}(w)^{-1}}{z-w}\right] \mathfrak C\ ,
}
where\footnote{%
In this computation we use $\epsilon ^{2}=\pi q/kp^2$, and $1-\epsilon ^{2}=-\pi/kq$.}
\begin{equation}
\phi _{\epsilon }(z)^{-1}=-\frac{k}{2\pi }pq\left[ \phi
(z)_\text{PCM}^{-1}+\epsilon ^{2}\phi (z)_\text{FR}^{-1}\right]
\label{interpolating twist}
\end{equation}%
and%
\begin{equation}
\phi (z)_\text{PCM}=\frac{1}{z^{2}}-1,\text{ \ \ }\phi (z)_\text{FR}=1,\text{ \ \ }%
\epsilon ^{2}=-\frac{1}{4\lambda }\left( 1-\lambda \right) ^{2}\ ,
\end{equation}%
are the twisting functions of the undeformed PCM and the Faddeev-Reshetikhin
(FR) model, and $\epsilon $ is the deformation parameter in the $\epsilon
^{2}<0$ branch.

The $\mathfrak r$/$\mathfrak s$ matrices satisfy a modified classical Yang-Baxter equation, mCYBE, a fact which
was proven explicitly in \cite{Sfetsos}. Notice that in the FR limit $%
\lambda \rightarrow 0$, we have 
\begin{equation}
\mathfrak s(z,w)=0
\end{equation}%
and the non-ultra-locality is completely removed. This is the celebrated
ultra-localization of the PCM introduced by hand in \cite%
{Faddeev-Reshetikhin}. Consequently, the $k$-PCM can be thought
as the Lagrangian implementation of the FR ultra-localization mechanism.

The values of the $s$ matrix at the poles are given by%
\begin{equation}
\mathfrak s(z_{\pm },z_{\pm })=\mp \frac{\pi }{k}\ ,\text{ \ \ \ \ \ }\mathfrak s(z_{\pm },z_{\mp
})=0\ .  \label{evaluations}
\end{equation}%
Using these values in the Maillet bracket (\ref{Lax-Lax PB}), we recover the
Kac-Moody current algebra (\ref{KM algebra}) in tensorial form
\begin{equation}
\big\{\overset{1}{\mathscr J}_{\pm}(x),\overset{2}{\mathscr J}_{\pm}(y)\big\} =\frac{1}{2}%
\big[ \mathfrak C,\overset{1}{\mathscr J}_{\pm}(x)-\overset2{\mathscr J}_{\pm}(x)\big] \delta
_{xy}\pm \frac{k}{2\pi }\mathfrak C\delta _{xy}^{\prime }\ .  \label{tensor KM}
\end{equation}%
The values of $\mathfrak r(z,w)$ at the same poles depends on how we take the limit
but are always proportional to the tensor Casimir and hence do not
contribute to the Poisson bracket. However, the values at different poles
vanish identically.

We end this section, by rewriting (\ref{Lax-Lax PB}) in a slightly different
way which will be useful when we consider the $k$-$F/G$ theories. By
defining%
\begin{equation}
\mathfrak r(z,w)=\frac{1}{2}\left( \mathfrak R(z,w)-\mathfrak R^{t }(z,w)\right)\ ,\quad\mathfrak s(z,w)=-\frac{1}{2}\left( \mathfrak R(z,w)+\mathfrak R^{t
}(z,w)\right)\ ,  \label{r/s matrices from R}
\end{equation}%
we can write the Maillet bracket as 
\begin{equation}
\big\{ \overset1{\mathscr L}(x,z),\overset2{\mathscr L}(y,w)\big\} =\big[ \mathfrak R,%
\overset1{\mathscr L}(x,z)\big] \delta _{xy}-\big[ \mathfrak R^{t },\overset2{\mathscr L}%
(y,w)\big] \delta _{xy}+\left( \mathfrak R+\mathfrak R^{t }\right) \delta
_{xy}^{\prime }\ ,  \label{Maillet 2}
\end{equation}%
where%
\begin{equation}
\mathfrak R(z,w)=-\frac{2\phi _{\epsilon }(w)^{-1}}{z-w}\mathfrak C\ .
\end{equation}

\noindent
{\bf\underline{ $k$-SSSM case}:} 

\noindent
Inspired by (\ref{interpolating twist}), this time we take the following inverse twisting function%
\begin{equation}
\varphi _{\epsilon }(z)^{-1}=c\left[ \varphi (z)_{\sigma }^{-1}+\epsilon
^{2}\varphi (z)_\text{FR}^{-1}\right]\ ,
\end{equation}%
where%
\begin{equation}
\varphi (z)_{\sigma }=\frac{4z^{2}}{(1-z^{2})^{2}}\ ,\text{ \ \ }\varphi
(z)_\text{FR}=1\ ,\text{ \ \ }\epsilon ^{2}=-\frac{1}{4\lambda}\left(
1-\lambda\right) ^{2}
\end{equation}%
and $c$ is a constant yet to be fixed by the evaluation of $\mathfrak s(z,w)$ at
the poles of $\varphi _{\epsilon }(z)$. 

The deformed twisting function is
\begin{equation}
c\,\varphi _{\epsilon }(z)=\frac{4}{z^{2}+2(2\text{\ }\epsilon ^{2}-1)+z^{-2}}
\end{equation}
which has four poles given by%
\begin{equation}
z_{+}^{2}=\lambda ^{-1},\text{ \ \ }z_{-}^{2}=\lambda.  \label{sigma poles}
\end{equation}%
The two poles corresponding to the positive branch of the square root are the evaluation points (\ref%
{evaluation points sigma}) used to extract the Lie-Poisson group charges out of
the monodromy matrix.\footnote{%
It would be interesting to find an interpretation for the negative branch poles and their relation with the monodromy matrix.} These
poles are important for establishing the quantum group
symmetry of the theory, as we shall see below.

The calculation we need to find $\mathfrak r$ and $\mathfrak s$ was already done in \cite{dialgebra} for the
branch $0\leq \epsilon ^{2}<1$ and also holds in the present case as the
compatibility of the Poisson brackets is independent of the sign of $%
\epsilon ^{2}$. The R-matrix corresponding to the Poisson
bracket in the form (\ref{Maillet 2}) is\footnote{In the following, we define $\mathfrak C^{(i,j)}=\sum_{T^a\in\mathfrak f^{(i)}}\sum_{T^b\in\mathfrak f^{(j)}}\eta_{ab}T^a\otimes T^b$ for $i,j\in\mathbb Z_2$. Only $\mathfrak C^{(0,0)}$ and $\mathfrak C^{(1,1)}$ are non-vanishing and $\mathfrak C=\mathfrak C^{(0,0)}+\mathfrak C^{(1,1)}$.}
\begin{equation}
\mathfrak R(z,w)=-\frac2{z^2-w^2}\left[\tsum\nolimits_{j=0,1}z^{j}w^{2-j}\mathfrak C^{(j,2-j)}\right]\varphi _{\epsilon }(w)^{-1}\ ,
\end{equation}%
from which follows the $\mathfrak r$/$\mathfrak s$ matrices in our present case%
\EQ{
\mathfrak r(z,w) &=-\frac{1}{z^{2}-w^{2}}\tsum\nolimits_{j=0,1}\left[
w^{j}z^{2-j}\varphi _{\epsilon }(z)^{-1}+z^{j}w^{2-j}\varphi _{\epsilon
}(w)^{-1}\right] \mathfrak C^{(j,2-j)}\ , \\
\mathfrak s(z,w) &=-\frac{1}{z^{2}-w^{2}}\tsum\nolimits_{j=0,1}\left[
w^{j}z^{2-j}\varphi _{\epsilon }(z)^{-1}-z^{j}w^{2-j}\varphi _{\epsilon
}(w)^{-1}\right] \mathfrak C^{(j,2-j)}\ .
}

Concentrating on the $\mathfrak s(z,w)$ matrix, after some simplification we
find it can be put in the form%
\begin{equation}
\mathfrak s(z,w)=-\frac{c}{4}\left[ z^{2}+w^{2}+2(2\text{\ }\epsilon ^{2}-1)%
\right] \mathfrak C^{(00)}+\frac{c}{4zw}\left[ 1-z^2w^2\right]
\mathfrak C^{(11)}\ .
\end{equation}%
Written in this way we can easily evaluate this function at the poles%
\begin{equation}
\mathfrak s(z_{\pm },z_{\pm })=\mp\frac{c}{4\lambda }(1-\lambda ^2)\mathfrak C\ ,%
\text{ \ \ }\mathfrak s(z_{\pm },z_{\mp })=0.
\end{equation}%
The normalization of the central term in the Kac-Moody algebra imposes the
condition 
\begin{equation}
-2\mathfrak s(z_{\pm },z_{\pm })=\pm \frac{2\pi }{k}\mathfrak C\,,
\end{equation}%
which fixes the so far unspecified constant $c$ to be%
\begin{equation}
c=\frac{4\pi \lambda}{k(1-\lambda ^{2})}\ .
\end{equation}%
In the limit $\lambda \rightarrow 0,$ we have 
\begin{equation}
\mathfrak s(z,w)=\frac{\pi }{k}\mathfrak C^{(00)}\ ,
\end{equation}%
from which follows that in the coset case the non-ultra-locality can be only
partially tamed~\cite{DMV 1}. As in the $k$-PCM case, from the Maillet bracket we recover the
Kac-Moody algebra in tensorial form.

\subsection{Expansion of the monodromy around the poles of the twisting
function}

We work in the de-compatified limit, in which the monodromy matrix
takes the general form\footnote{%
Take $g_{s}=-i$ and $A=-{\mathscr L}$ in eq. (20) of \cite{Wilson} and
re-organize eqs. (22).}%
\begin{equation}
m=\text{Pexp} \left[ -\dint_{a}^{b}dx\,{\mathscr L}(x)\right] =\exp \left[
\dsum\nolimits_{n=1}^{\infty }F_{n}\right] ,  \label{expansion monodromy}
\end{equation}%
where for the first three terms we get 
\begin{eqnarray*}
F_{1} &=&-\dint_{a}^{b}dx\,{\mathscr L}(x)\ , \\
F_{2} &=&\frac{1}{4}\dint_{a}^{b}dx\,dy\,\left[ {\mathscr L}(x),{\mathscr L}(y)%
\right] \epsilon _{xy}\ , \\
F_{3} &=&-\frac{1}{12}\dint_{a}^{b}dx\,dy\,dz\,\left[ {\mathscr L}(x),\left[ 
{\mathscr L}(y),{\mathscr L}(z)\right] \right] \epsilon _{xy}\epsilon _{yz}\ .
\end{eqnarray*}%
The limit $a$ and $b$ will subsequently be taken to $\pm\infty$, respectively.
Expanding ${\mathscr L}(x)$ around a chosen point in the complex plane and
reorganizing the series in powers of the spectral parameter $z$, not to be confused with the integration variable $z$, we find the
corresponding non-local charges associated to that particular point.

Let us focus first on the $k$-PCM theory. When expanded
around the poles $z_{\pm }=\mp p/q,$ we find the following expression for
the spatial Lax operator 
\begin{equation}
\mathscr{L(}x\mathcal{)=}-j_{(0)\pm }(x)-j_{(1)\pm }(x)\widehat{z}_{\pm
}-j_{(2)\pm }(x)\widehat{z}_{\pm }^{2}-j_{(3)\pm }(x)\widehat{z}_{\pm }^{3}+%
\mathcal{O(}\widehat{z}_{\pm }^{4}\mathcal{)}\ ,  \label{Lax expanded}
\end{equation}%
where $\widehat{z}_{\pm }=z-z_{\pm }$ and 
\EQ{
j_{(0)\pm } &=\pm \frac{1}{4\lambda }(1+\lambda )\left[ (1-\lambda
)I_{0}\pm (1+\lambda )I_{1}\right] =\mp \frac{2\pi }{k}\mathscr J_{\pm }=-%
\mathscr L(z_{\pm })\ , \\
j_{(1)\pm } &=\frac{1}{8\lambda ^{2}}(1-\lambda )^{2}\left[ (1+\lambda
^{2})I_{0}\pm (1-\lambda ^{2})I_{1}\right]\ , \\
j_{(2)\pm } &=\pm \frac{1}{16\lambda ^{3}}(1-\lambda )^{3}\left[ (1+\lambda
^{3})I_{0}\pm (1-\lambda ^{3})I_{1}\right]\ , \\
j_{(3)\pm } &=\frac{1}{32\lambda ^{4}}(1-\lambda )^{4}\left[ (1+\lambda
^{4})I_{0}\pm (1-\lambda ^{4})I_{1}\right]\ .
\label{PCMexpansion}
}

The expansion for the Lax operator around the poles \eqref{Lax expanded} starts with terms of order $\widehat z_\pm^{0}$. These terms can be removed by judicious gauge transformations,
\begin{equation}
m'(z)=u(b)m(z)u(a)^{-1},\text{ \ \ }{\mathscr L}^{\prime }=u(%
\mathscr L+\partial_{x})u^{-1}\ ,
\end{equation}%
where the gauge transformation takes the form
\begin{equation}
u(x,x_{0},z_{\pm })=\text{Pexp} \left[ -\dint\nolimits_{x_{0}}^{x}dy\,\mathscr L(%
z_{\pm }\mathcal{)}\right] \equiv u(x)_{\pm }  \label{wave function}
\end{equation}
around each pole, respectively.
In this way, (\ref{expansion monodromy}) can be written as follows 
\begin{equation}
m(z)=u(b)_{\pm }\text{Pexp} \left[ -\dint_{a}^{b}dx\,{\mathscr L}'(x)_{\pm }%
\right] u(a)_{\pm }^{-1},
\end{equation}%
with a transformed Lax operator%
\begin{equation}
{{\mathscr L}}'(x)_{\pm }=-j'_{(1)\pm }(x)\widehat{z}_{\pm }-%
j'_{(2)\pm }(x)\widehat{z}_{\pm }^{2}-j'_{(3)\pm }(x)%
\widehat{z}_{\pm }^{3}+\mathcal{O(}\widehat{z}_{\pm }^{4}\mathcal{)}
\end{equation}%
defined in terms of the non-local dressed currents%
\begin{equation}
j'_{(n)\pm }(x)=u(x)_{\pm }^{-1}j_{(n)\pm }(x)u(x)_{\pm },\text{ \
\ }n\geq 1\ .
\end{equation}

Taking $x_{0}=a=-\infty $ and $b=\infty $, the expansion of
the monodromy matrix around the poles $z_{\pm }$ now takes the form
\begin{equation}
m(z)=m(z_{\pm })\exp \left[ \dsum\nolimits_{n=1}^{\infty }q_{(n)\pm }%
\widehat{z}_{\pm }^{n}\right]\ .  \label{pole expansion}
\end{equation}
As a result of the gauge transformation, all the charges $q_{(n)\pm }$ are non-local and the first three take the explicit form\EQ{
q_{(1)\pm } &=\dint_{-\infty }^{\infty }dx\,j'_{(1)\pm }(x)\ , \\
q_{(2)\pm } &=\frac{1}{4}\dint\nolimits_{-\infty }^{\infty }dxdy\left[ 
j'_{(1)\pm }(x),j'_{(1)\pm }(y)\right] \epsilon
_{xy}+\dint\nolimits_{-\infty }^{\infty }dx\,j'_{(2)\pm}(x)\ , \\
q_{(3)\pm } &=\frac{1}{12}\dint\nolimits_{-\infty }^{\infty }dxdydz\left[ 
j'_{(1)\pm }(x),\left[ j'_{(1)\pm }(y),j'_{(1)\pm
}(z)\right] \right] \epsilon _{xy}\epsilon _{yz}\\ &\qquad\qquad+\frac{1}{2}%
\dint\nolimits_{-\infty }^{\infty }dxdy\,\left[ j'_{(1)\pm }(x),%
j'_{(2)\pm }(y)\right] \epsilon _{xy} +\dint\nolimits_{-\infty }^{\infty }dx\,j'_{(3)\pm }(x)\ .
}%
Notice that under 
$\lambda \rightarrow \lambda ^{-1}$, which corresponds to the transformation~\eqref{parity}, we do indeed have $q_{(n)\pm
}\rightarrow q_{(n)\mp }$ as claimed. This follows from our earlier observation that we can combine the $F_{L}$
group action with~\eqref{parity} to generate $F_{R}$ and vice-versa.
If we evaluate these charges at the, for us, unphysical value of the coupling constant $\lambda ^{\prime }=-1$, then the two poles coalesce at $z_\pm=0$ and the expansion of the monodromy produces charges that coincide 
with the first three Yangian charges found in \cite{Sfetsos}
\begin{equation}
q_{(1)\pm }(\lambda ^{\prime })\equiv Q_{(0)},\text{ \ \ }q_{(2)\pm
}(\lambda ^{\prime })\equiv Q_{(1)},\text{ \ \ }q_{(3)\pm }(\lambda ^{\prime
})\equiv Q_{(2)}.
\end{equation}%

On the one hand, the dominant term for the expansion around $z=0$ is the local
Noether charge $Q_{(0)}$ corresponding to the global vector action of $F$ and given by \eqref{vector charge}.
The Poisson algebra generated by $Q_{(0)}$ is isomorphic to $\mathfrak{f}$ and the higher charges $Q_{(n)}$, $n>0$ enhance this Poisson algebra to a
Yangian, see \cite{Sfetsos}. 

On the other hand, the dominant terms for the
expansions around $z_{\pm }=\mp p/q$ are the non-Abelian moments (\ref%
{non-Abelian moments}) associated to the global Poisson-Lie group action $%
F_{L}\times F_{R}$. They will be shown to generate the finite-dimensional quantum group $%
U_{q}(\mathfrak f)$ after quantization. It is then natural to
expect that higher charges enhance $U_{q}(\mathfrak f)$ to an
affine quantum symmetry to match the symmetries of the  S-matrices that we have conjectured to describe these theories in section \ref{s2}. However, explicit computations of the algebra of higher charges is complicated by the non-trivial dressing factors 
$u(x)_{\pm }$, hence we expect to report about this problem in the near future. Fortunately, some evidence that this enhancement may occur in our case comes from a different kind of deformation of the $\SU(2)$ PCM \cite{Yoshida1,Yoshida2,Yoshida3}, in which a classical $q$-deformed Poisson algebra of symmetries induced by the squashing is studied for this particular case. What suggests this is that, technically, the situation in those references is essentially the same as ours, namely, the generators of the $q$-deformed symmetry group are extracted from the expansion of the monodromy matrix at the poles of the twisting function.

For the $k$-SSSM theories, a similar story holds. We have exactly
the same expansion as given by \eqref{Lax expanded} but with different expressions for the current components%
\EQ{
j_{(0)\pm } &=I_{1}^{(0)}-\frac{1}{2}\left( \lambda ^{\pm 1/2}-\lambda
^{\mp 1/2}\right) I_{0}^{(1)}+\frac{1}{2}\left( \lambda ^{\pm 1/2}+\lambda
^{\mp 1/2}\right) I_{1}^{(1)}\\
&
=\mp \frac{2\pi }{k}\mathcal{J}_{\pm }=-%
{\mathscr L}(z_{\pm }\mathcal{)}, \\
j_{(1)\pm } &=\frac{1}{2}(1+\lambda ^{\pm 1})I_{0}^{(1)}+\frac{1}{2}%
(1-\lambda ^{\pm 1})I_{1}^{(1)}, \\
j_{(2)\pm } &=-\frac{1}{2}\lambda ^{\pm 3/2}\left(
I_{0}^{(1)}-I_{1}^{(1)}\right) , \\
j_{(3)\pm } &=\frac{1}{2}\lambda ^{\pm 2}\left(
I_{0}^{(1)}-I_{1}^{(1)}\right),
}
where $\widehat{z}_{\pm }=z-\lambda ^{\mp 1/2}.$ An intuitive way to seek for Yangian symmetries is to find a value of $\lambda$ for which $j_{(0)\pm }=0$, c.f the first equation of \eqref{PCMexpansion} with $\lambda=-1$ in the $k$-PCM case. Notice, however, that such a solution does not exist for the $k$-SSSM theories, hence we do not expect any Yangian symmetries in this case. This is in agreement with the absence of global Noether charges as notice above around \eqref{noNoether}. 

Then, the expansion of the monodromy matrix around the poles takes the rather universal
form \eqref{pole expansion}. The model dependence shows up in the higher charges $q_{(n)\pm },$ $%
n\geq 1.$ In the $k$-PCM they are all Lie-algebra valued while in the $k$-$F/G$ they are all coset valued.

We now proceed to quantize the monodromies $m(z_{\pm })$ and find their quantum algebra.

\section{Quantum lattice current algebra}\label{s4}

In order to quantize a integrable field theory, the technical difficulties in dealing with monodromy matrix are usual addressed by introducing a lattice regulator which does not break the integrability of the theory. Unfortunately in the current setting, our theories have a non-ultra-local Poisson bracket and it is well known  
 \cite{Maillet} that, even at the classical level, this precludes a direct approach to computing the Poisson bracket of
the monodromy matrix for any value of the spectral parameter $z$. 

In the quantum theory, this means that the lattice current
algebra for these field theories for general values of the spectral parameter $z$ is not even known to exist. 
Fortunately, we can make progress by working directly at the poles $z=z_\pm$ of the twist function. Our approach in this section draws mainly on the formalism established in \cite{Alekseev,Unraveling,Faddeev}. Like in these references, our calculations will be restricted to ${\mathfrak f}={\mathfrak su}(n)$.

\subsection{The lattice current algebra at the poles}

Let us recall that the equations of motion of the $k$-deformed theories can be re-cast in Lax form as the 
zero curvature condition \eqref{zcc}. This form manifests an
infinite-dimensional group of gauge transformations  generated by the group elements $\Gamma (x,z)$%
\begin{equation}
{\mathscr L}_{\pm }^{\prime }(z)=\Gamma (z){\mathscr L}_{\pm }(z)\Gamma
(z)^{-1}-\partial _{\pm }\Gamma (z)\Gamma (z)^{-1}\ .
\end{equation}%
From (\ref{evaluation}) we conclude that, when evaluated at the poles $z_{\pm
}$, this group becomes the ordinary loop group $g_{\pm }(x)=\Gamma (x,z_{\pm
})$ of gauge transformations of the Kac-Moody currents%
\begin{equation}
\mathscr J_{\pm }^{\prime }(x)=g_{\pm }(x)\mathscr J_{\pm }(x)g_{\pm
}(x)^{-1}\pm \frac{k}{2\pi }\partial _{x}g_{\pm }(x)g_{\pm }(x)^{-1}\ ,
\label{class gauge transf}
\end{equation}%
which are symmetries of the Kac-Moody algebra \eqref{tensor KM}.

Before introducing a lattice regularization of the theory we can infer, via
a semiclassical analysis, that the $k$-deformation parameter of the quantum
group generated by $m(z_{\pm })$ is indeed a root of unity. This information
can be extracted from the Maillet representation (\ref{Lax-Lax PB})
evaluated at the poles. Introducing the parallel transport
\begin{equation}
u(x,x_{0})=\text{Pexp} \left[ \dint\nolimits_{x_{0}}^{x}dy\,{\mathscr L}(y)\right]\ ,
\end{equation}%
which obeys 
\begin{equation}
\partial _{x}u(x,x_{0})={\mathscr L}(x)u(x,x_{0}),\text{ \ \ }\partial
_{x_{0}}u(x,x_{0})=\mathcal{-}u(x,x_{0}){\mathscr L}(x_{0})\ ,
\end{equation}%
and then following Maillet \cite{Maillet}, we find in
our case 
\begin{equation}
\big\{\overset1u(x,x_{0}),\overset2u(y,y_{0})\big\}
=\dint\nolimits_{x_{0}}^{x}dz\dint\nolimits_{y_{0}}^{y}dz^{\prime
}\,\overset1u(x,z)\overset2u(y,z^{\prime })\big\{\overset1{\mathscr L}(z),\overset2{\mathscr L}(z^{\prime })\big\} \overset1u(z,x_{0})\overset2u(z^{\prime },y_{0})\ .
\end{equation}%

In the above, for brevity, we have not indicated the spectral parameter dependence but we should keep in mind that $u$ depend on it. In fact, we
are only interested in the result above when it is evaluated at the poles 
$z_{\pm }$ in which ${\mathscr L}(z)$, as in \eqref{evaluation}, picks out the Kac-Moody currents. In
this case we find that%
\begin{equation}
\big\{\overset1u(x,x_{0}),\overset2u(y,y_{0})\big\}
=s\,\overset1u(x,x_{0})\overset2u(y,y_{0})\left( \epsilon _{xy}+\epsilon
_{x_{0}y_{0}}\right)\mathfrak C\ ,  \label{classical exchange}
\end{equation}%
where $s$ is given by%
\begin{equation}
\mathfrak s(z_{\pm },z_{\pm })=s(z_{\pm })\mathfrak C\ ,\text{ \ \ \ \ \ }s(z_{\pm
})=\mp \frac{\pi }{k}\ ,
\end{equation}%
and $\epsilon _{xy}$ is the sign function and in what follows we will drop the $\pm $ indices for simplicity.\footnote{In what follows we fix the
ambiguity on the lower points of integrations by setting $x_{0}=y_{0}$,
which is the normalization condition reproduced in the classical continuum
limit of the lattice algebra regularization to be considered below. We will drop explicit dependence on $x_0$ from the notation.}

The expression (\ref{classical exchange}) is the classical exchange algebra
we want to quantize semi-classically. We want to emphasize that at the moment we are proceeding in a rather ad-hoc fashion in order to provide some inspiration for the lattice algebra defined below.
By exploiting the classical
commutativity of the $u$'s and using the quantization rule\footnote{%
This kind of na\"\i ve replacement is subtle when group-like quantities have quadratic Poisson brackets
but this subtlety is not relevant in the semi-classical limit.} $-i\left\{ \ast
,\ast \right\} \rightarrow\left[ \ast ,\ast \right] $, we get%
\begin{eqnarray*}
\overset1u(x)\overset2u(y) &=&\overset2u(y)\overset1u(x)\left( 1-is\,\mathfrak C\,\epsilon
_{xy}+\cdots\right)\\
&=&\overset2u(y)\overset1u(x)\mathfrak R_{xy}\ ,
\end{eqnarray*}%
where the ellipsis denote sub-leading quantum contributions. From this we
infer that, roughly speaking, we should expect something like%
\begin{equation}
\mathfrak R_{xy}\thicksim q^{\mathfrak C\epsilon _{xy}}\ ,\quad q\thicksim e^{-is}\ ,\quad q^k\thicksim 1\ ,  \label{semi-classical input}
\end{equation}%
which is an interesting expression as it relates Maillet's $s$ matrix,
the poles and the $k$-deformation parameter. Notice that the effect of the
tranformation~\eqref{parity} is to map $q\rightarrow q^{-1}.$

The calculation above should not be taken too
seriously, we only use it only as a source of inspiration for the type of quantum algebra we should expect. In particular, the  Poisson brackets (\ref{classical exchange}) do not obey the
Jacobi identity. Indeed, we find%
\begin{equation}
\text{Jacobi}=-s^{2}\,\overset1u(x)\overset2u(y)\overset3u(z)\left[ \epsilon _{xy}\epsilon _{yz}+\epsilon
_{xz}\epsilon _{zy}+\epsilon _{zx}\epsilon _{xy}\right] f^{abc}T^{a}\otimes
T^{b}\otimes T^{c},
\end{equation}%
which never vanishes for distinct $x,y,z$.

After the na\"\i ve semi-classical approach, we are now ready to introduce the lattice algebra. The first thing is to
discretize the spatial circle $S^{1}$ in terms of $N$ points with coordinates $x_{n}=n\Delta$, $
n=1,\ldots,N$ mod $N$ in the order $x_{1}<x_{2}<\cdots<x_{N-1}<x_{N}$. Then, we
define the transport matrix%
\begin{equation}
t(x_{n+1},x_{n})=\text{Pexp} \left[ \dint_{x_{n}}^{x_{n+1}}dx\,\mathcal{L(}x\mathcal{%
)}\right]
\end{equation}%
between sites. Evaluating it at the poles $z_{\pm }$ defines for us the
lattice variables 
\begin{equation}
J_{n}^{L}=\text{Pexp} \left[ \frac{2\pi }{k}\dint_{x_{n}}^{x_{n+1}}dx\mathscr J%
_{+}(x)\right] ,\text{ \ \ }J_{n}^{R}=\text{Pexp} \left[ -\frac{2\pi }{k}%
\dint_{x_{n}}^{x_{n+1}}dx\mathscr J_{-}(x)\right]\ .
\end{equation}%
Choosing a particular ordering, we can define two monodromy
matrices on $S^{1}$,
\begin{equation}
M^{L}=J_{N}^{L}\cdots J_{1}^{L}\text{, \ \ \ \ \ }M^{R}=J_{1}^{R}\cdots J_{N}^{R}\ .
\label{monodromies}
\end{equation}%
However, in order to keep things simpler we will concentrate on the pole $%
z_{+}$ only, drop the index $L$ and set $g_{+}(x)=g(x).$

The quantum lattice Kac-Moody algebra at the pole $z_{+}$ is then defined by the
following algebra
\begin{equation}
\overset{1}{J}_{n+1}\mathfrak R^{-}\overset{2}{J}_{n}=\overset{2}{J}_{n}\overset{1}{J}%
_{n+1}\ ,\text{ \ \ \ \ \ }\overset{1}{J}_{n}\overset{2}{J}_{n}=\mathfrak R^{+}\overset{2%
}{J}_{n}\overset{1}{J}_{n}\mathfrak R^{-}\ ,  \label{Left LKMA}
\end{equation}%
where $\mathfrak R^{+}=\mathfrak R^{t}$ (transpose) and $\mathfrak R^{-}=\mathfrak R^{-1}$ (inverse). 

Now, the classical continuum limit is achieved by writing%
\begin{equation}
J_{n}\sim 1+\frac{2\pi }{k}\Delta \mathscr J_{+}(x),\text{ \ \ }\mathfrak R\sim
1+i\gamma \overline{\mathfrak r}\ ,
\end{equation}%
where $\mathfrak R$ and $\overline{\mathfrak r}$ are solutions of the quantum/classical YBE such
that 
\begin{equation}
\overline{\mathfrak r}+\overline{\mathfrak r}^{t}=2\mathfrak C\ ,  \label{reg. r-matrix}
\end{equation}%
where $\Delta =1/N\rightarrow 0$ is the lattice spacing and $x=n/N$. The first
step is to find the classical limit of (\ref{Left LKMA}). We find
\begin{equation}
\frac{1}{\gamma }\big\{ \overset{1}{J}_{m},\overset{2}{J}_{n}\big\} =%
\overset{1}{J}_{m}\overset{2}{J}_{n}\overline{\mathfrak r}\delta _{mn}-\overline{\mathfrak r}%
^t\overset{1}{J}_{m}\overset{2}{J}_{n}\delta _{mn}-\overset{1}{J}_{m}%
\overline{\mathfrak r}\overset{2}{J}_{n}\delta _{m,n+1}+\overset{2}{J}_{n}\overline{\mathfrak r}%
^t\overset{1}{J}_{m}\delta _{m+1,n}\ .
\end{equation}%
The next step is to take the continuum limit. We find that $\gamma =-s$
and then (\ref{tensor KM}) follows. The algebra (\ref{Left LKMA}) underlies the quantization of the
monodromy matrix $m(z_{+})$.

Recall that classically we have%
\begin{equation}
\mathscr J_{+}(x)=\frac{k}{2\pi }\partial _{x}u(x)u(x)^{-1}\ ,
\label{KM classical input}
\end{equation}%
where $u(x)$ is the ``wave-function''%
\begin{equation}
u(x)=\text{Pexp} \left[ \frac{2\pi }{k}\dint_{x_{0}}^{x}dx\mathscr J_{+}(x)\right]
\ .
\end{equation}%
The gauge transformations (\ref{class gauge transf}) acting on $\mathscr J%
_{+}(x)$ are now equivalent to the action of the loop group $u(x)\rightarrow
g(x)u(x)$ on the $u^{\prime }s$. On the lattice, this classical expressions
become\footnote{%
For any value of $n,$ the $u_{n}$ all start with the term $J_{1}$, which is
the normalization condition $x_{0}=y_{0}$ used before in the semi-classical
analysis.}%
\begin{equation}
u_{n}=J_{n}\cdots J_{1},\text{ \ \ }u_{n}=J_{n}u_{n-1},\text{ \ \ }%
u_{n}\rightarrow g_{n}u_{n},\text{ \ \ \ \ \ \ }J_{n}\rightarrow
g_{n}J_{n}g_{n-1}^{-1}.
\end{equation}%
The wave functions also obey the following quantum exchange algebra relations%
\EQ{
\overset{1}{u}_{m}\overset{2}{u}_{n} &=\overset{2}{u}_{n}\overset{1}{u}%
_{m}\mathfrak R^{+}\ ,\text{ \ \ \ }m>n\ , \\
\overset{1}{u}_{m}\overset{2}{u}_{n} &=\overset{2}{u}_{n}\overset{1}{u}%
_{m}\mathfrak R^{-}\ ,\text{ \ \ \ }m<n\ , \\
\overset{1}{u}_{n}\overset{2}{u}_{n} &=\mathfrak R^{+}\overset{2}{u}_{n}\overset{1}{u}%
_{n}\mathfrak R^{-}\ .  
 \label{lattice exchange} 
}
The last relation means that the classical gauge loop group is promoted to a
quantum gauge group with algebra \cite{Unraveling}%
\begin{equation}
\overset{1}{g}_{n}\overset{2}{g}_{n}\mathfrak R^{+}=\mathfrak R^{+}\overset{2}{g}_{n}\overset{1}{%
g}_{n}\,\text{ \ \ \ \ \ }\overset{1}{g}_{m}\overset{2}{g}_{n}=\overset{2}{g}%
_{n}\overset{1}{g}_{m}\ ,\text{ \ \ }m\neq n\ .
\end{equation}%
This quantum group leaves the Kac-Moody lattice algebra (\ref{Left LKMA})
invariant in the same way its classical counterpart leaves invariant the
usual Kac-Moody algebra.

Writing $\overline{\mathfrak{r}}=-\mathfrak{r+C,}$ with $\mathfrak{r}^{t}=-%
\mathfrak{r}$ antisymmetric which explicitly obeys the constraint (\ref{reg. r-matrix}), we can find the
corrected version of the Poisson bracket (\ref{classical exchange}) for the wave-function $u(x)$
(in the normalization $x_{0}=y_{0})$ directly from the classical/continuum
limit of the first two equations of the exchange algebra (\ref{lattice exchange}), namely,%
\begin{equation*}
\left\{ \overset{1}{u}(x),\overset{2}{u}(y)\right\} =s\overset{1}{u}(x)%
\overset{2}{u}(y)\left[ \mathfrak{r+C}\epsilon _{xy}\right] .
\end{equation*}%
This Poisson bracket do obey the Jacobi identity provided the antisymmetric $%
\mathfrak{r}$-matrix is a solution of the mCYBE of the split-type \cite%
{Vicedo:2015pna}. See the discussion around eq. (5.9) of that paper. This
decomposition of $\overline{\mathfrak{r}}$ is consistent with the
requirement that it is a solution of the CYBE.

The quantum monodromy matrix is given by $M=u^{N}$ and satisfy the quadratic
algebra
\begin{equation}
\overset{1}{M}\mathfrak R^{-}\overset{2}{M}(\mathfrak R^{-})^{-1}=\mathfrak R^{+}\overset{2}{M}(\mathfrak R^{+})^{-1}%
\overset{1}{M}\ .  \label{quantum G star}
\end{equation}%
When we consider the following quantum factorization of the monodromy matrix%
\begin{equation}
M=M_{-}^{-1}M_{+}\ ,  \label{Q RH problem}
\end{equation}%
the $RTT$ quadratic relations given by its factorized components, namely,%
\begin{equation}
\mathfrak R^{+}\overset{1}{M}_{\pm }\overset{2}{M}_{\pm }=\overset{2}{M}_{\pm }\overset%
{1}{M}_{\pm }\mathfrak R^{+}\ ,\text{ \ \ \ \ \ }\mathfrak R^{+}\overset{1}{M}_{-}\overset{2}{M}%
_{+}=\overset{2}{M}_{+}\overset{1}{M}_{-}\mathfrak R^{+}\ ,
\label{factor quantum G star}
\end{equation}%
are compatible with (\ref{quantum G star}). Notice that (\ref{factor
quantum G star}) is the analogue of eq.~(13) of \cite{Faddeev} in the case
of $sl(2)$, which is another definition of the associated quantum group $%
Fun_{q}(SL(2)^{\ast })$. The quantum monodromy matrix $M(z_{+})$ then
becomes an element of $F_{q}$. The Poisson-Lie symmetry found in the
classical $k$-deformed theory is then promoted to a quantum group
symmetry.

The continuum limit of (\ref{lattice exchange}) is%
\begin{equation}
\overset{1}{u}(x)\overset{2}{u}(y)=\overset{2}{u}(y)\overset{1}{u}(x)\mathfrak R_{xy}\ ,%
\text{ \ \ \ \ \ }\mathfrak R_{xy}=\mathfrak F\,q^{\mathfrak P\epsilon _{xy}}\,\mathfrak F^{t-1}\ ,
\label{quantum exchange}
\end{equation}%
where $\mathfrak P$ is the permutation operator and $\mathfrak F$ an invertible matrix. This way
of factorizing the usual $R$ matrix comes from the theory of quasi-Hopf
algebras introduced by Drinfeld.\footnote{For further details we refer, for example, to section 2 of 
\cite{Sanchez} in which the main properties of Drinfeld twists are
reviewed.} It was shown in \cite{Faddeev}, that this exchange algebra
underlies the quantization of the Kac-Moody algebra. Indeed, the quantum version of
the classical currents (\ref{KM classical input}), namely, the quantum
operators 
\begin{equation}
j(x)=\partial _{x}u(x)\cdot u(x)^{-1}
\end{equation}%
obey the following algebra%
\begin{equation}
\big[ \overset{1}{j}(x),\overset{2}{j}(y)\big] =-\ln q\big[
\mathfrak P,\overset{1}{j}(x)-\overset{2}{j}(y)\big] \delta _{xy}-2\ln
q(1+a)\mathfrak P\delta _{xy}^{\prime }\ ,  \label{Q KM}
\end{equation}%
where%
\begin{equation}
a=-\frac{n\ln q}{i\pi }
\end{equation}%
and $n$ is the size of the matrices used to represent the currents, e.g, $n$
for the fundamental representation of $\mathfrak{sl}(n)$, which corresponds to the present case. Taking the ansatz for
the deforming parameter 
\begin{equation}
q=e^{i\pi /(k+n)}\ ,
\end{equation}%
and re-scaling the quantum currents (take $k\rightarrow k+n$ in (\ref{KM
classical input})) it was shown in \cite{Faddeev} that the rescaled currents
obey the quantum Kac-Moody algebra. Notice that in the semi-classical limit
we recover our previous rough result with $q=e^{-is(z_{+})}$. Using the duality
between the Faddeev-Reshetikhin-Takhtajan $(F_{q})$ and the Drinfeld-Jimbo $(U_{q}(\mathfrak{f}))$
realizations of quantum groups, we infer that the deformed sigma
models considered so far are $q$-deformed with deformation parameter
being a root of unity.

A comment is in order. From the second equation in (\ref{r/s matrices from R}) and (\ref{reg. r-matrix}) one
would naively conclude that%
\begin{equation}
\mathfrak R(z_{+},z_{+})=\frac{\pi }{k}\overline{\mathfrak r}\ ,
\end{equation}%
but the following observation shows this is not the case. The $z\rightarrow z_{+},w\rightarrow z_{+}$ limit is
ambiguous as can be seen from 
\begin{equation}
\mathfrak{R(}z,z_{+}\mathfrak{)=}0,\text{ \ \ \ \ }\underset{w\rightarrow
z_{+}}{\lim }\text{\ }\mathfrak{R(}z_{+},w\mathfrak{)}\sim \mathfrak{C}
\end{equation}%
and even if we ignore this and work with the non-zero answer, this identification would imply
that the classical limits of (\ref{quantum G star}) and (\ref{factor quantum G star}) are trivial, which is inconsistent. Then, the poles of the twisting function are to be treated as punctures in the complex plane $z$ with a complete separate analysis. In this respect, the quantum $\mathfrak{R}$-matrix and its
classical counterpart $\overline{\mathfrak{r}}$ used in (\ref{Left LKMA}) are artifacts of our quantization scheme
not related to the $\mathfrak{r/s}$ operators in the Maillet
bracket at all. In the classical/continuum limit, quantities should be
independent of the quantization procedure but if, for instance, we consider the symmetry
algebras (\ref{quantum G star}) and (\ref{factor quantum G star}) they would depend on $\overline{%
\mathfrak{r}}$ in this limit, hence the quantized monodromy matrix should remain quantum. Notice that this does not apply to its associated KM algebra because it is independent of $\overline{\mathfrak{r}}$ in the classical/continuum limit, as claimed in \cite{Faddeev}.

It is noteworthy that it is the central charge $k$ of the Kac-Moody algebra instead of
parameter $\lambda $ that sets the deformation of the symmetry group. This is, perhaps,
surprising as one might have expected $q$ to depend on the deformation parameter $%
\lambda $ as in the $\eta $-deformation \cite{eta 1}. Presumably the reason
lies in the fact that the Kac-Moody currents (\ref{KM currents}) obey the
Kac-Moody algebra no matter what the deformation parameter $\lambda$ is and hence their
monodromies obey a $\lambda$-independent algebra. In this respect, the $k$-deformations ($-\infty <\epsilon ^{2}\leq 0$) are quite different in nature from the $%
\eta $-deformation ($\epsilon ^{2}\in \lbrack 0,1)$). However, there is
still the possibility that somehow higher spin charges appearing in the $z$
expansion of the monodromy matrix around the poles, e.g (\ref{pole
expansion}), introduce a $\lambda $ dependence in a way yet to be understood.


\section{Conclusions}\label{s5}

In this paper, we have conjectured the form of the S-matrix theories that should describe the $k$-deformed principal chiral models and symmetric space sigma models. All of them exhibit affine quantum group symmetries with a deformation parameter $q$ that is a root of unity. 
Then, making use of the results of~\cite{Faddeev}, we have provided evidence that the quantum group symmetry appears in a direct quantization that preserves integrability 
in the form of a finite-dimensional quantum group symmetry, at least for the theories corresponding to $\mathfrak{f}=\mathfrak{su}(n)$.

However, there are some important details that remain to be understood. For instance, the proposed S-matrices exhibit a symmetry under the affine quantum group while it seems difficult to see how to extend the symmetry of the lattice quantization in this way.
A proper study of the higher spin non-local charges that appear in the expansion of the monodromy matrix at the poles of the twisting function is certainly interesting from the quantum group theory point of view. In this regard, it would be important to understand the role of these charges in relation to the conjectured affine quantum group symmetry of the S-matrices for the deformed theories.

It is also not immediately clear how to generalize the approach of \cite{Faddeev} to the cases of theories involving the other classical Lie groups.  Even more ambitious, would be the generalization to semi-symmetric space sigma models involving Grassmann fields and Lie superalgebras that are needed to apply these ideas to the superstring. In any case, leaving out the technical details, it is worth recalling that the existence of the finite-dimensional quantum group symmetry follows simply from eq.~\eqref{evaluation}. Namely, from the observation that the spatial component of the Lax operator of these theories evaluated at specific values of the spectral parameter, which are 
poles of the $\lambda $-deformed twisting function, is a Kac-Moody current. Remarkably, the same is true for the $k$-deformation of the Green-Schwarz $AdS_5\times S^5$ sigma model (see eq.~(3.14) of\cite{HMS fermionic}) and, thus, we expect that a similar construction will reveal a finite-dimensional quantum group symmetry with a deformation parameter $q$ that is a root of unity also in this case.

\section*{Acknowledgements}\
\noindent The authors would like to thank to Benoit Vicedo for very valuable
communications at an early stage in this work. DMS is very
grateful to Alessandro Torrielli for very pleasant and enlightening
discussions around quantum groups and to the Surrey University for
hospitality during his visits.

TJH is supported in part by the STFC grant ST/G000506/1. 
JLM is supported in part by MINECO (FPA2011-22594 and FPA2014-52218-P), the Spanish Consolider-
Ingenio 2010 Programme CPAN (CSD2007-00042), Xunta de Galicia (GRC2013-024),
and FEDER. DMS is supported by the FAPESP grant 2012/09180-9.

\end{document}